# Surface-aware Mesh Texture Synthesis with Pre-trained 2D CNNs

Áron Samuel Kovács , Pedro Hermosilla, and Renata G. Raidou

TU Wien, Austria

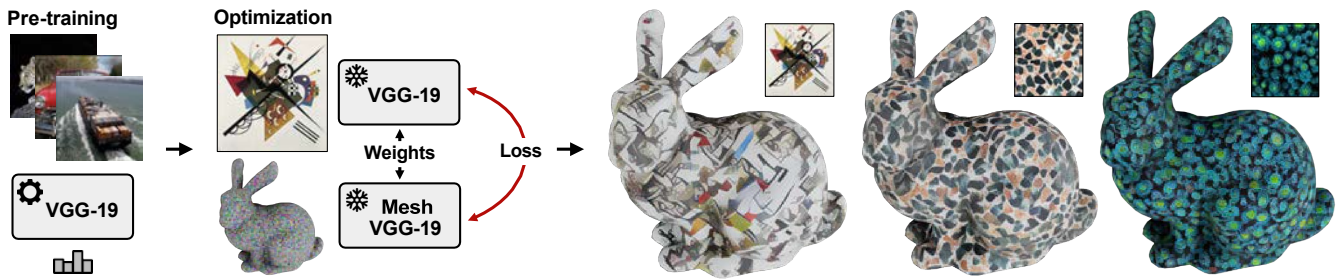

**Figure 1:** *Our mesh texture synthesis algorithm employs two neural networks with the same architecture: the first one is a conventional 2D Convolutional Neural Network (CNN) designed for images, while the second operates directly on the tangent space of a mesh. The shared architecture allows us to use the weights of the 2D CNN—pre-trained on thousands of natural images—on the CNN operating on the mesh. Consequently, with the parameters of the networks frozen, we optimize the content of a mesh texture to match the content of a specified image.*

**Abstract**

*Mesh texture synthesis is a key component in the automatic generation of 3D content. Existing learning-based methods have drawbacks—either by disregarding the shape manifold during texture generation or by requiring a large number of different views to mitigate occlusion-related inconsistencies. In this paper, we present a novel surface-aware approach for mesh texture synthesis that overcomes these drawbacks by leveraging the pre-trained weights of 2D Convolutional Neural Networks (CNNs) with the same architecture, but with convolutions designed for 3D meshes. Our proposed network keeps track of the oriented patches surrounding each texel, enabling seamless texture synthesis and retaining local similarity to classical 2D convolutions with square kernels. Our approach allows us to synthesize textures that account for the geometric content of mesh surfaces, eliminating discontinuities and achieving comparable quality to 2D image synthesis algorithms. We compare our approach with state-of-the-art methods where, through qualitative and quantitative evaluations, we demonstrate that our approach is more effective for a variety of meshes and styles, while also producing visually appealing and consistent textures on meshes.*

## 1. Introduction

In *computer graphics*, textures refer to two-dimensional images applied to the surfaces of 3D meshes, by mapping the pixels of the former onto the vertices or polygons of the latter. During mapping, we determine which pixel in the texture maps to a mesh surface point in consideration—ultimately simulating a diverse range of surface properties, such as color, reflectivity, transparency, and more. This process is fundamentally oriented towards creating sophisticated and realistic visual effects, leveraging highly intricate texture images while preserving rendering efficiency. Conversely, in *computer vision*, textures refer to visual patterns or structures intrinsic to an image. These patterns or structures are identified through the spatial arrangement of pixel intensities, color distributions, and other visual features. Texture analysis, entailing the extraction and characterization of these patterns, focuses on revealing the underlying structure of an image and understanding its content.

Generating high-quality textures for 3D meshes is a manual and tedious process, due to the inherent disparity between the 2D nature of textures and the 3D shape to which they are eventually projected. Consequently, the automatic generation of textures for 3D meshes, known as *mesh texture synthesis* has emerged as a crucial technique for the rapid and controllable creation of 3D content. Situated at the intersection of computer graphics and computer vision, mesh texture synthesis aims to generate textures for 3D meshes that ex-hibit visual coherence, meaningfulness, and realism. To accomplish this, mesh texture synthesis takes into account both the underlying geometry and topology of a 3D mesh, as well as the underlying structure or pattern present within the texture image. Mesh texture synthesis may enable, thus, the creation of visually compelling and contextually appropriate texture representations for 3D meshes.

Several works have addressed the problem of controllable texture synthesis from a *2D perspective*. These methods operate by taking a 2D image exemplar as input and generating random variations thereof. Traditional methods either define mechanical processes for generating such variations [EL99, EF01, WLKT09] or employ a parametric texture model [Jul62, PS00, SF95]. In recent years, Convolutional Neural Networks (CNNs) have also been trained on natural images to generate variations of a given exemplar [GEB15b, JAFF16]. Although these methods produce visually appealing results, they are not explicitly designed for mesh textures since they are unaware of the eventual geometric context.

Alternative approaches have tackled the same problem from a *3D perspective*. One such approach is solid texture synthesis, which aims to generate a texture in 3D space based on a 2D exemplar. In this context, colors are associated with specific positions within a bounded or unbounded 3D volume [HMR20, GRGH19a, KFCO*07]. However, these methods are designed to generate high-frequency and abstract textures, such as marble, and do not consider the lower-dimensional manifold of 3D meshes during the generation process. Another line of research employs 2D CNNs in conjunction with a differentiable renderer to optimize the resulting mesh texture from a given 2D exemplar across multiple viewpoints [HJN22a, MPSO18a]. These methods heavily rely on sampling a large number of views, which can be computationally demanding and may lead to occlusion-related inconsistencies.

In this paper, we propose a novel surface-aware mesh texture synthesis method that mitigates the drawbacks observed in prior research efforts. Our methodology leverages the pre-trained weights of a 2D CNN to another CNN with an identical architecture—but with convolutions operating *directly on the tangent space* of a mesh. This design enables the utilization of pre-existing weights from the 2D network, trained on an extensive corpus of natural images. Moreover, it facilitates the comparison of Gram matrices between the two networks, as we define a loss function between a 2D texture and a mesh texture directly. In a comparative evaluation, we demonstrate that our approach generates visually appealing mesh textures from a large variety of exemplars while respecting the geometric context of the mesh. Our implementation is publicly available in our repository [KHR24].

## 2. Related Work

Our work builds upon advances in geometric deep learning, specifically focused on texture synthesis (Sec. 2.1). These advances utilize a generalization of the convolution operator to process meshes with a neural network (Sec. 2.2).

### 2.1. Image Synthesis

**2D Approaches.** Textures are generated using two main strategies. The first strategy entails resampling either pixels [EL99, WL00] or entire patches [EF01, KSE*03] of the original texture. A complete review of non-parametric resampling techniques is provided by Wei et al. [WLKT09]. Such techniques are capable of producing natural textures very efficiently—yet, they do not provide an actual texture model. The second strategy is to explicitly define a parametric texture model [Jul62, PS00, SF95]. Although these approaches are effective for a wide range of textures, they are not sufficient in representing natural textures [GEB15b].

Image and texture generation approaches can be categorized into two main groups within the context of our research. The first category involves the generation of new textures with high-level features, derived from an exemplar. The second category further tries to preserve the content of an additional input image. Often, both categories share similar methodologies, requiring only minor modifications before being used interchangeably. In both cases, the goal is to reuse features from the exemplar, which can be constrained by features in the content image, if desired.

The foundational work in neural style transfer has been laid by Gatys et al. [GEB15a], where the technique was developed to transfer artistic styles. This concept has been since repurposed for synthesizing a texture by using a noise image as the content input [GEB15b]—thus, not preserving any content. Notable advancements have been made in both style transfer and texture synthesis. These include the integration of Generative Adversarial Networks (GANs) [JBV17], the design of specialized networks tailored for features of varying scales [ZGW*22], and the training of neural networks that explicitly minimize the objective function [CS16].

**3D Approaches.** The problem of synthesizing textures for 3D objects can be approached from different angles. Point clouds techniques [CWN19, CWNN20] assign color values to points in space. However, point clouds face inherent challenges in unambiguously expressing surfaces, due to the absence of explicit connections between individual points. In many cases, though, this is desired to prevent bleeding of features across surfaces that just happen to be nearby. A recent survey by Guo et al. [GWH*20] provides a review of recent progress in deep learning for point clouds.

When the style image aims to represent a material that can be represented with a volume, such as wood or marble, solid texture approaches come into play [HMR20, GRGH19a, ZGW*22, CW10, KFCO*07]. Solid texture methods directly assign values, such as color or density, to every point in 3D space that can be sampled onto a mesh. Hence, during the synthesis part of the pipeline, these methods are not aware of the mesh's surface. Not all materials or styles can be represented with a solid texture that exhibits a high degree of symmetry, as features are correlated based on Euclidean distance and do not consider the separation on the surface.

To address these shortcomings, render-based approaches adopt a process of rendering the given mesh and subsequently matching the style of the projected parts [KUH17, MPSO18a, HJN22a]. Such methods require rendering the object from multiple viewpoints, thereby, introducing challenges when dealing with objects of high complexity, with multiple holes, or overlapping wire-like features, where achieving visibility of all surfaces becomes difficult. Nonetheless, when the camera is suitably positioned, the projection process can effectively capture the separation of surfaces by empty space. Consequently, these methods can break correlations solely based on Euclidean distance, at least during rendering.

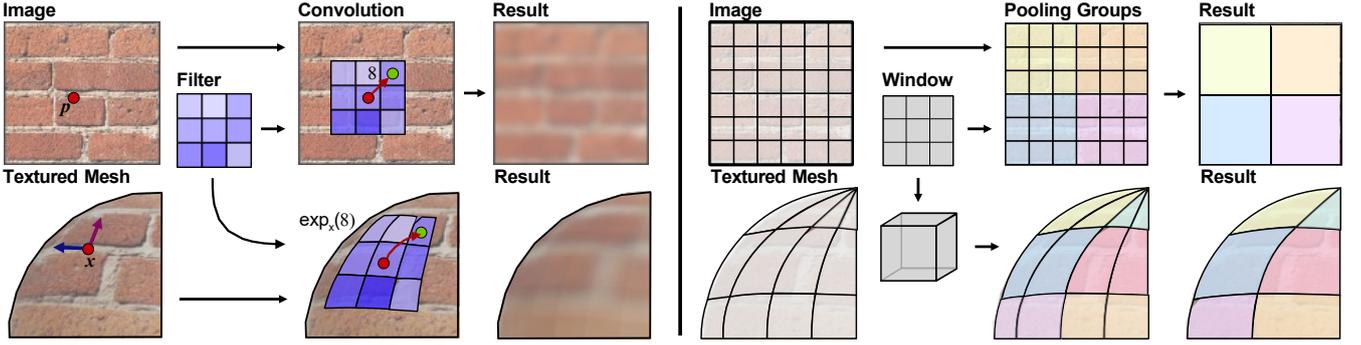

**Figure 2:** *The convolution and pooling operation, as redefined within the context of our work. Left: The convolution is applied to the input data (image vs. textured mesh) to filter the available information and produce a feature map. However, for the textured mesh, we modify the neighborhood sampling to account for the mesh topology. Right: During the pooling, we define a sliding 3D window that selects texels to aggregate based on their geodesic path (indicated with the colors).*

Recently, text-based approaches have also gained popularity as methods for mesh texture generation, since they allow to specify the content of the texture with simple prompts [MBOL*21, KXBP22, RMA*23]. These methods also rely on 2D projections from multiple views to generate a set of renderings that are then fed to a pre-trained CLIP model [RKH*21]. Therefore, they suffer from the same shortcomings as render-based methods.

### 2.2. Convolutional Neural Networks for 3D Meshes

The design of neural network architectures for meshes is an active area of research. While vertex-based approaches that represent the mesh as a graph exist [YHSG17, VBV18, FLFM18, HHF*19], they cannot easily capture data not solely associated with individual vertices or faces. Other approaches have defined the convolution operations directly on the Riemannian manifold. Among these, two approaches have been considered: Using a diffusive approach related to heat diffusion on surfaces [SACO22, WNEH22, MBM*17, MBBV15], or equivariant convolutions on surfaces designed to address the rotation ambiguity problem of the tangent plane [PO18, YLP*20, MKK21]. Lastly, texture-based approaches have emerged as a viable solution, as textures enable the definition of data on a finer level than just the pure geometry of a mesh [HZY*19, LLZ*19, GWY*21]. All these approaches have been designed for a supervised setup where the downstream task is performed directly on the mesh surface. Therefore, these methods are not a viable solution when the parameters of another neural network, trained on 2D image data, need to be re-used.

### 3. Our Approach

Our method, inspired by previous works in the field of 2D texture synthesis [GEB15b], introduces a novel spatial invariant parametric texture model. This model is built upon a hierarchical CNN that has been pre-trained on the task of object recognition in natural images. In contrast to previous works, we extend this concept to 3D meshes by generalizing the building blocks of the 2D network architecture to operate on the tangent space of a 3D mesh (Fig. 1). This generalization involves redefining two of the main building blocks of the network: the *convolution operation* and the *pooling operation*. Fig. 2 illustrates schematically these operations.

Whilst the convolution operation applies filters on the image to detect patterns (Sec. 3.1), the pooling operation reduces the image size to increase the receptive field of the subsequent filters and reduce the computational burden (Sec. 3.2). By redefining these operations to operate on the surface of a mesh, we can directly leverage the weights of a pre-trained 2D network, while maintaining the rest of the architecture intact. To this end, our method optimizes 3D mesh convolution and pooling by precomputing geodesic neighborhoods and pooling groups with linear time complexity relative to texel counts and triangle numbers (Sec. 3.3). Consequently, we generate textures directly on the surface of a 3D mesh (Sec. 3.4). In the following subsections, we describe in detail each of these operations and the optimization process of our proposed algorithm.

### 3.1. Convolution

**2D Definition.** Given an input feature map $F_i \in \mathbb{R}^{W \times H}$, the standard 2D discrete convolution computes an output feature map $F_o \in \mathbb{R}^{W \times H}$. The new feature value for each pixel $p \in \mathbb{N}^2$ is the weighted sum of input features from pixels in its neighborhood $N$. The weights of this sum are defined by a kernel function $g(\delta)$ that takes the relative position $\delta \in \mathbb{Z}^2$ of the neighboring pixel w.r.t. $p$:

$$F_o(p) = \sum_{\delta \in N} F_i(p + \delta) g(\delta) \qquad (1)$$

The neighborhood $N$ is defined as a set of vectors covering a square area of $k \times k$, which is usually implemented as a matrix of the same $k \times k$ shape. In this definition, we omit the third dimension of the feature maps describing multiple channels for simplicity.

**3D Mesh Definition.** To re-use the weights of a 2D convolution in our mesh convolutions, we use the same definition of convolution as in Eq. 1. However, we modify the neighborhood sampling to account for the mesh topology. We assume that our 3D meshes are 2D Riemannian manifolds $M$ without boundary embedded in 3D space. Therefore, we can make use of exponential maps to describe the sampling in Eq. 1 for each point $x$ in our manifold:

$$F_o(x) = \sum_{\delta \in N} F_i(\exp_x(\delta)) g(\delta) \qquad (2)$$

where $\delta \in T_x M$, being $T_x M$ the tangent plane at point $x$. The exponential map $\exp_x(\delta)$ will follow the geodesic path $\gamma_\delta(1)$ and retrieve a neighboring point $y$ in that direction. Therefore, the square area defined by the neighborhood $N$ in Eq. 1 is now covering a square area in the manifold locally around $x$. Note that this definition of convolution is directly inspired by Masci et al. [MBBV15], but we use square neighborhoods instead of circular ones.

**Feature Maps.** While in Eq. 1 the feature maps $F$ were images, now in Eq. 2 $F$ denotes scalar-valued functions on the manifold $F : M \rightarrow \mathbb{R}$. Different representations can be used for $F$. In this paper, we choose to represent $F$ as mesh textures. Given the set of points $x$ on $M$ and a 2D texture map $F$, we define a bijective map $t : M \rightarrow [0, 1]^2$ that maps points on $M$ to points in $F$. Our convolution then becomes:

$$F_o(p) = \sum_{\delta \in N} F_i(p') g(\delta) \qquad (3)$$
$$p' = t(\exp_{t^{-1}(p)}(\delta))$$

Since $F_i$ is now a discretized 2D feature map, we need to adjust the length of the geodesic path $\gamma_\delta(1)$ based on the distance between texels in our feature map. Our vector $\delta$, therefore, becomes $\delta' = \delta s$, where $s$ is the size of a texel in world space.

To sample our texture map $F$ at continuous positions, we need to define an interpolation function. Nearest neighbor and bilinear interpolations are fast to execute and implemented by hardware. In our experiments, we used bilinear interpolation in the first layer and nearest neighbor interpolations in the remaining layers of the network, having empirically confirmed that it yields better results. We use the *xatlas* [You22] library to generate the UV mappings, though this choice was primarily motivated by the ease of use and any method that tries to preserve areas should work comparably.

**Tangent Frame.** Our new definition of convolutions requires a local reference frame defined at each point $x$. For simplicity, given the normal $N$ at point $x$, we compute the tangent vectors $T$ and $B$ with the Gram-Schmidt process using $N$ and a random vector $w$. However, this process is not defined when $N$ and $w$ are parallel. In such cases, we use a different vector $w'$, which is perpendicular to $w$. Note that $w$ and $w'$ are the same for all reference frames. Our algorithm is not bound to this method for computing the tangent frames and other methods that generate smoother tangent frames could be used, e.g., the approach by Fisher et al. [FSDH07].

### 3.2. Pooling

**2D Definition.** The pooling operations used for images usually define a window of size $k \times k$ that is overlaid over the image and a pooling operation that aggregates the values inside the window. For simplicity, in this paper, we will consider non-overlapping windows, i.e., each pixel is not used by more than one $k \times k$ window.

**3D Mesh Definition.** To define a similar pooling operation for meshes, we need to define a sliding window that selects texels to aggregate. We use a simple—yet effective—algorithm by defining a 3D voxelization of the space and aggregating texels within a voxel that are connected through a geodesic path within this voxel. The voxel grid of each pooling layer is defined as $k^n \cdot s$, where $n$ is the index of the pooling layer within the network starting from 1, $s$ is the size of the texel in world space, and $k$ the size of the window. Note that this algorithm could generate several pooling groups inside the same voxel. Also note that this algorithm will generate pooling groups of different sizes.

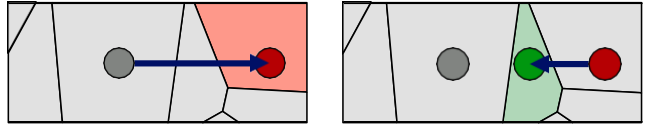

**Figure 3:** *We employ a mechanism to avoid overshooting from a texel at position p (in grey) to a distant disconnected texel (in red). Therefore, we correctly determine the adjacent texel (in green).*

**Texel Graph.** To accelerate the pooling operation, we build an undirected graph $E$ of the texels of our texture at each level in the network. An edge $e_{ij}$ on this graph connects texels $i$ and $j$ if they are adjacent, i.e., if there is a boundary connecting the two texels. This allows for fast queries of connected components during the determination of pooling groups. Our pooling operation generates pooling groups of varying sizes and, thus, texels of varying sizes in deeper layers of the network. This might cause our exponential maps to overshoot and select a distant texel that is not connected to the central texel at $p$, for which the convolution is being computed. We depict this in Fig. 3. To avoid such cases, we use our graph $E$ to determine if the neighboring point is adjacent to our central texel. If not, we follow $\gamma_{\delta'}(1)$ backward, until we find a point along $\gamma_{\delta'}(1)$ belonging to a texel adjacent to our central texel at $p$.

### 3.3. Precomputation

Unlike regular 2D convolution and pooling, where the position of each neighbor is trivial to compute, in 3D mesh convolution and pooling it is not sufficient to sample surrounding points in UV space. However, the local geodesic neighborhood of each texel can be computed in advance, assuming that a given mesh is not being deformed during the training process. For the convolution, our implementation precomputes the local neighborhood of each texel, storing the position and bilinear factor for each sample in GPU buffers. The time complexity of this process is at worst linear w.r.t. the number of triangles and the number of texels, as the computation of the geodesic paths consists primarily of crossing from one face to another. During training, our method only depends on the number of texels in each layer and should scale linearly.

Similarly, we precompute the grouping into pooling groups which we then again store on the GPU. This has a linear time complexity w.r.t. the number of texels as this mostly involves grouping them into voxels and then computing the connected components using the aforementioned texel graph. As a UV unwrapping may not fully utilize the whole UV space, we work only with the texels that are used and store only their features. This approach enables us to ignore unused texture parts—thus, reducing the needed memory.

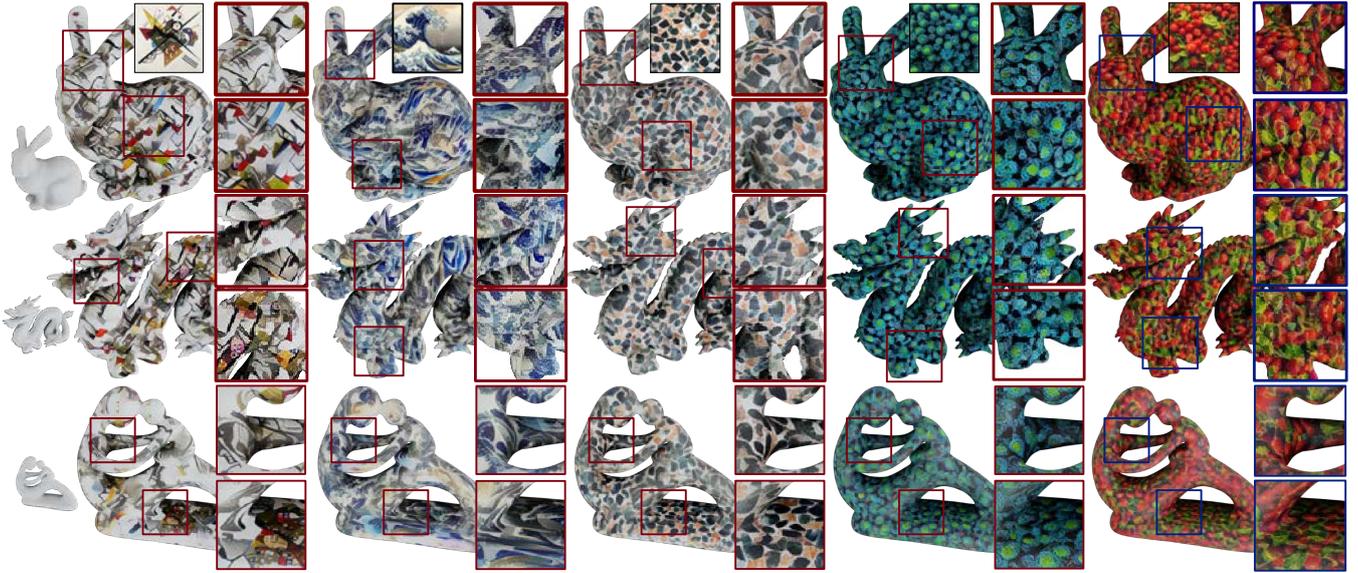

**Figure 4:** *Our method applied on three meshes (the bunny and the dragon from the Stanford 3D scanning repository, and the Mother and Child by Brian Weston (CC BY-SA)) and five textures with a diverse set of stimuli.*

### 3.4. Texture Synthesis

Our algorithm makes use of two CNNs with the same architecture. We used the VGG-19 architecture [SZ14], which is one of the most commonly used for image synthesis and style transfer for 2D images [GEB15a, GEB15b, LW16]. The choice between the architecture variants of VGG was based on the documented advantages of VGG-19 over VGG-16 with regard to its increased discriminative power, improved representational capacity of complex images, and better feature extraction due to the deeper architecture. Whilst one of our networks is designed to work on 2D images, the other one uses convolution and pooling operations for 3D meshes, as schematically depicted in Fig. 1.

We start with a VGG-19 network, pre-trained for the task of image classification on ImageNet1000 [RDS*15]. Subsequently, we use the pre-trained weights to initialize the second network that operates on 3D meshes. Once the weights of both networks are defined, they are frozen before the texture generation. We initialize the mesh texture with random noise values drawn from a uniform distribution on the range [0, 0.2]. The choice has been motivated by the findings of previous work [GEB15b]. Then, we execute the 2D network with the 2D image example as input, and the mesh network with the random texture as input. For each layer of the network, the Gram matrices are computed [GEB15b]. The final loss is defined as the mean square error between the Gram matrices of all layers in the network, where all of them are assigned the same weight. In this way, we measure the discrepancy between the Gram matrices, reflecting the difference in style representation between the generated outcome and the target. The gradients of this loss are back-propagated through the mesh network until the input mesh texture and its values are optimized. After several optimization steps, the final mesh texture is generated. A schematic depiction of our approach is provided in Fig. 1.

### 4. Results and Evaluation

In this section, we present an analysis of the results generated with our method (Sec. 4.1). We also provide a comparison to other indicative state-of-the-art approaches (Sec. 4.2). All synthesis results are generated using our tool written in a combination of Python, Rust, TensorFlow, and CUDA, and running on a desktop machine equipped with an AMD Ryzen 9 3900X with 128 GB RAM and an NVIDIA GeForce RTX 3080 Ti. We use the Adam optimizer with a rather large learning rate of 0.1, which is halved after 200 iterations and again after 400 iterations. The optimization process stops after 500 iterations. Depending on the complexity of the mesh and the fraction of the utilized UV space, the precomputation part takes 2–5 minutes and the optimization process takes approximately 50 minutes for textures of size $1024 \times 1024$. More detailed measurements are included in Tables 1 and 2.

### 4.1. Visual Quality of our Approach

To demonstrate the versatility of our approach, we prepared a selection of testing sequences with a variety of meshes and textures as input data. These include meshes with different topologies and at different levels of detail (i.e., coarseness), and texture exemplars with different stimuli and resolutions.

**Results with Different Meshes and Textures.** Fig. 4 depicts the outcomes of our approach applied to the bunny and the dragon from the Stanford 3D scanning repository, as well as the Mother and Child mesh, for five textures with a diverse set of stimuli—ranging from high-frequency textures [PS00] to isotropic materials [GRGH19a], and also artistic styles [GEB15b]. More results with additional meshes and textures are included in Fig. 10. The results indicate that our approach can infer a reasonable mesh texture from the 2D exemplar while preserving the features learned along the mesh, for a large variety of textures. Our method suc-cessfully reproduces the underlying structure of the different exemplars, while capturing their colors and variations, and also following the mesh geometry. For instance, when employing textures with *Kandinsky's on White II* or *Hokusai's The Great Wave off Kanagawa*, we observe that colors and patterns (fine-grained and coarse) are preserved, while the geometrical context of the surface is also respected—even for meshes of different topologies or higher genus, such as the dragon or the Mother and Child mesh in the last two rows of Fig. 4. For the plant texture, we notice a few bright spots in all meshes (e.g., at the back foot of the bunny in the fourth column of Fig. 4). This might be due to rapid changes in the tangent field in the surrounding area. This may be causing the mesh network with VGG-19 weights to lose local context to such an extent that the extracted local features cannot be properly matched with the provided style. For the plant and the radishes textures (fourth and fifth column of Fig. 4), we also notice that the shapes are slightly more elongated as opposed to the rounder structures in the exemplar. This might be due to the different sizes and shapes of the pooling groups.

**Results at Different Levels of Detail for the Meshes and Different Texture Resolutions.** We also experimented with different resolutions for the textures and different levels of details, i.e., coarseness, for the meshes. In Fig. 5 (a), we depict one example with the armadillo from the Stanford 3D scanning repository at two different levels of detail (2,124 and 212,574 polys) and the *Kandinsky* texture at two different resolutions ($128 \times 128$ and $256 \times 256$). As anticipated, a higher texture resolution and a higher number of polygons both influence the outcome. Although the texture resolution qualitatively seems to be the most influential factor, it is noteworthy that using a finer mesh also contributes to fewer artifacts, e.g., on the chest of the armadillo in Fig. 5 (a).

**Impact of Weight Initialization.** Moreover, we evaluate the effect of the pre-training process in 2D. Fig. 5 (b) offers a comparison of the results of our optimization process when using the weights of the 2D pre-trained network against those of a neural network with randomly initialized weights. We see that, without the prior knowledge acquired during pre-training, the algorithm is not able to generate a plausible mesh texture (left). On the other hand, when the pre-trained weights are used, the generated mesh texture matches the patterns of the original image (right).

**Impact of the Overshooting Correction.** We finally evaluate the effect of our correction that shortens geodesic paths to avoid overshooting as shown in Fig. 3. Fig. 5 (c) shows that without this correction, artifacts can be seen in certain parts of the texture such as the bunny tail. The pre-trained network does not expect the overshooting to occur and, therefore, skipping over to more distant unconnected parts of the texture may cause it to extract incorrect features.

### 4.2. Comparison to the State of the Art

We compare our approach to selected approaches from the state of the art—namely, the approaches of Gatys et al. [GEB15b], Gutierrez et al. [GRGH19a], Mordvintsev et al. [MPSO18a], and Höllein et al. [HJN22a] (Fig. 6). The work of Gatys et al. is a traditional 2D image texture synthesis neural approach. Gutierrez et al. propose a solid texture approach, where a volume is defined and sliced so that the style of the slices matches a given style image. Afterward, the volume is sampled at the object-space texel positions. Conversely, the approaches of Mordvintsev et al. and Höllein et al. are both render-based. The former renders a textured mesh and then matches the Gram matrices with a style image, while the latter introduces additional viewing angle and depth corrections. These works showcase different perspectives and applications, ranging from traditional 2D image style transfer to volumetric texture synthesis and style transfer in rendering 3D scenes. However, they all focus on generating or transforming images in a way that captures and transfers specific visual styles or textures. The approach of Gatys et al. is only used as a baseline for the 2D case, while the other three are used for the 3D scenario comparison.

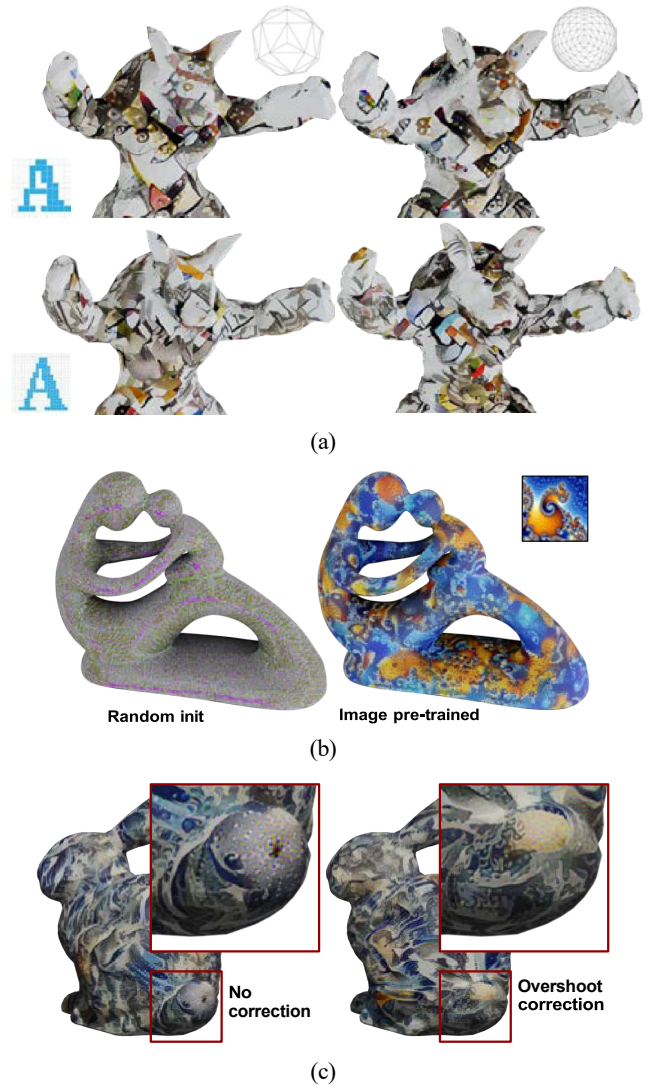

(a)

(b)

(c)

**Figure 5:** *Ablation study: (a) Our method applied to the armadillo from the Stanford 3D scanning repository (at two different levels of detail: 2,124 and 212,574 polys) and the Kandinsky texture (at two different resolutions: $128 \times 128$ and $256 \times 256$). (b) Our method applied to the Mother and Child by Brian Weston (CC BY-SA) once with randomly initialized weights (left) and once with pre-trained VGG-19 weights. (c) Our method applied to the bunny from the Stanford 3D scanning repository without (left) and with (right) the overshooting correction.*

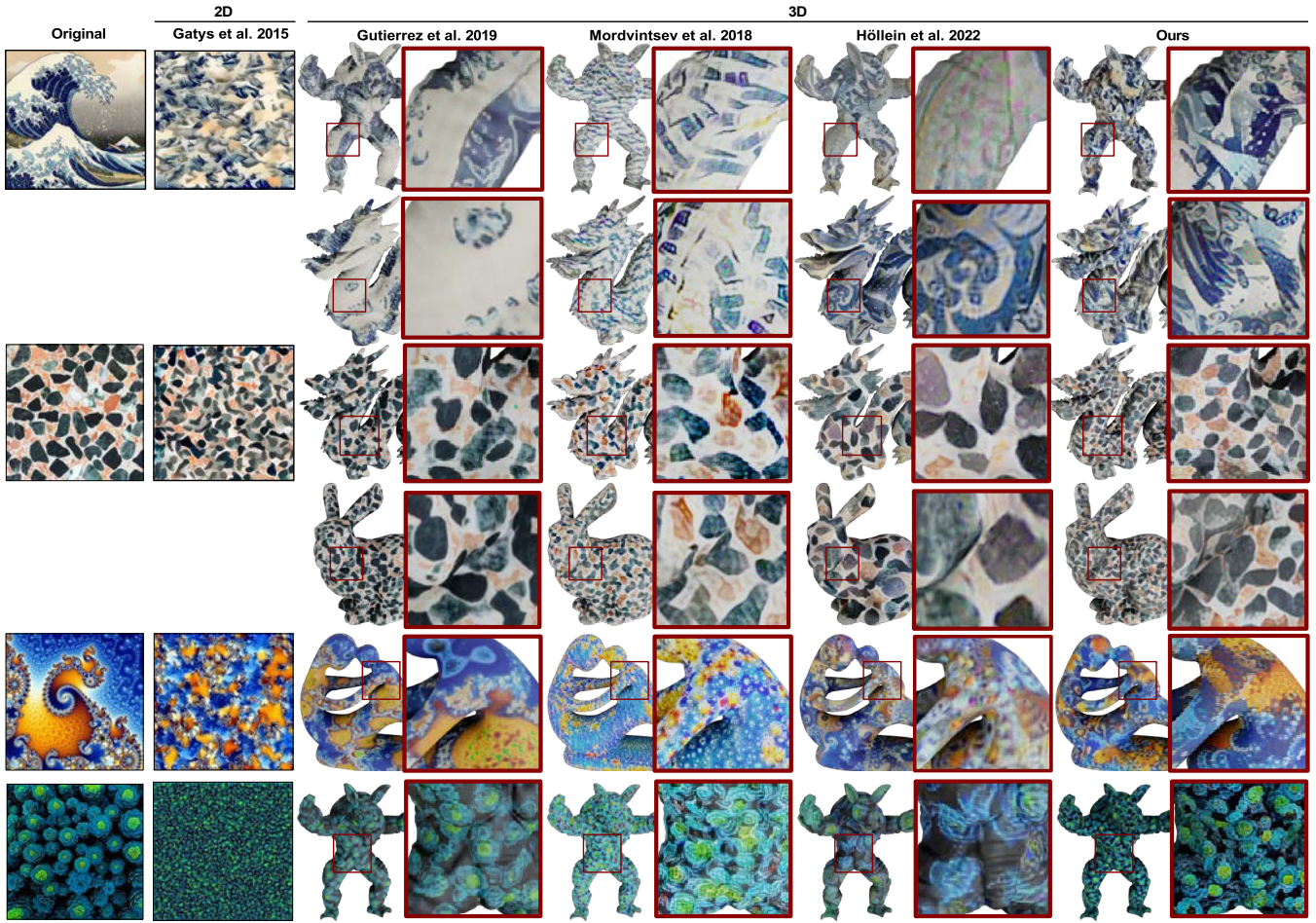

**Figure 6:** *Results of our approach compared to those by Gatys et al. [GEB15b], Gutierrez et al. [GRGH19a], Mordvintsev et al. [MPSO18a], and Höllein et al. [HJN22a] for four meshes (armadillo, dragon, and bunny from the the Stanford 3D scanning repository, and the Mother and Child by Brian Weston (CC BY-SA)) and four textures.*

Our selection has also been motivated by the availability of open-source implementations [GEB15c, MPSO18b, GRGH19b, HJN22b]. To maintain uniformity across methodologies, we deliberately chose to employ a VGG-19 network consistently in our study. As Gutierrez et al. and Höllein et al. already utilize VGG-19 in their approaches, we adapted our methodology to incorporate this network. It is worth noting that we had to transform the original implementation of Mordvintsev et al., which initially employed an Inception v1 network [SVI*16]. The quality of results is contingent on the chosen network, and we aimed to employ a consistent network across all approaches. Therefore, we had to integrate a VGG-19 network instead, ensuring a cohesive and comparable evaluation framework across all approaches. Still, the usage of VGG-19 is not fundamental for our approach and that is also the case for the neural networks chosen by the approaches we compare against. All approaches use these 2D networks just as feature extractors and the contribution of all these works is the method to bridge the gap between 3D and 2D (rendering or slicing) and/or corrections to account for this projection. We additionally note that the approach of Höllein et al. is primarily meant to stylize scenes. To adapt it to synthesize textures for individual meshes, we had to place cameras uniformly around each mesh and used the output of each hidden layer to compute the Gram matrices. Finally, we used the same UV unwrapping for all approaches.

**Visual Comparison.** For the comparison to the state of the art, we generated comparable cases for all approaches. The comparison outcomes are shown in Fig. 6 for *four meshes* (the armadillo, the dragon, and the bunny from the the Stanford 3D scanning repository, as well as the Mother and Child mesh) and *four textures* (*The Great Wave*, a marble texture, a texture depicting the Mandelbrot set, and a texture with succulent plants from GitHub). As anticipated, the approach of Gatys et al. is capable of producing natural 2D textures both for *The Great Wave* and the isotropic marble material texture. We observe, however, a few artifacts—for instance, at the bottom right corner of *The Great Wave* texture and also at the left edge of the marble texture. These might be due to the proximity to the edges of the generated image, where zero-padding does not match the distribution of values in the rest of the image, and the network extracts unsuitable features. Finally, for the Mandelbrot texture and the plants texture, the method is also able to capture the structure and patterns of the original texture, but at a different scale due to the different texture resolutions used in the original texture and the generated one.

Subsequently, we move on to the comparison with the approaches of Gutierrez et al. [GRGH19a], Mordvintsev et al. [MPSO18a], and Höllein et al. [HJN22a]. The approach of Gutierrez et al. works particularly well for isotropic materials, such as the marble texture. This is anticipated, because isotropic materials exhibit a high degree of symmetry, and as such any two slices of the stylized volume should resemble each other. With regard to *The Great Wave*, we see that the surface of the mesh is not well-respected (e.g., at the thigh of the armadillo in Fig. 6). Here, we have harsh artifacts, which can be observed as darker stripes on the white part of the texture. The same kind of artifacts can also be seen when employing the plants texture (e.g., at the belly of the armadillo in Fig. 6). Moreover, the colors and the sharpness are compromised to a degree, while for the Mandelbrot texture bright artifacts appear (e.g., at the child's leg in Fig. 6).

Conversely, the approach of Mordvintsev et al. does not work as expected. A potential reason for that is the use of VGG-19, as opposed to the original Inception v1 network. Although Inception networks are anticipated to yield superior results, comparing them to our employed VGG-19 would be unfair. Therefore, we leave this investigation as a point for future work—potentially, within a wider ablation study. The presence of blurry artifacts in render-based baselines could be attributed to the selection of camera viewpoints, as uniform placement might result in certain patches being inadequately viewed from optimal, i.e., orthogonal, perspectives. For *The Great Wave*, the outcomes of the approach of Mordvintsev et al. are vaguely resembling the exemplar, where only the colors are preserved. Additionally, some artifacts resembling speckle noise are visible, which are also evident in the images generated with the approach of Gatys et al. [GEB15b]. This can be mitigated by blurring or by including a term in the loss function that penalizes noise. The same artifacts are also present in the marble case, although the overall visual quality of the result is better with this texture. For the Mandelbrot and the plants texture, we notice severe artifacts that compromise both the represented structures and the saturation of the texture.

Finally, the performance of the approach of Höllein et al. seems to depend on the employed mesh. Notably, for *The Great Wave*, the performance with the armadillo mesh is suboptimal, while the dragon has only a few speckle-like artifacts. The marble texture works well for both employed mesh specimens, while the performance with the Mandelbrot and plants textures is insufficient. For the Mandelbrot case, the colors are desaturated and large artifacts are evident, while large streak artifacts appear in the plants texture also. An additional reason for the artifacts in the results obtained with Höllein et al.'s implementation could be due to *mipmapping*. This is a technique where a texture is progressively downsampled to increase rendering speed and reduce aliasing artifacts. In neural approaches, mipmapping can be used as a pooling operation [LLZ*19] or as a method for working with coarser features [HJN22a]. However, this approach may introduce errors. In some cases, the UV mapping may generate islands adjacent to each other in the UV space—yet, far apart in a geodesic sense. These islands merge during mipmapping, causing the gradient to flow through incorrectly merged pixels, and potentially resulting in erroneous correlations between regions or introducing undesirable noise. This drawback is counteracted in our implementation by pooling based on the geodesic space (instead of the UV space). We, therefore, suggest that techniques utilizing mipmapping should also use adequate island margins during training, or should consider alternative methods for pooling or representing coarse features.

**User Study.** To evaluate our method, we conducted an informal, online user study with 30 participants, where we used several of the cases shown in Fig. 4 and 10. We presented each participant with the produced outcomes of our approach and the approaches of Gutierrez et al., Mordvintsev et al., and Höllein et al. together with the respective texture exemplars. Without disclosing any information about any of the approaches, we interviewed the participants to gain some qualitative feedback about the outputs. Namely, we asked them to rank the four approaches concerning their similarity to the provided texture exemplar. For each of the generated results, we also asked the study participants to rate on a 1–5 Likert scale their visual appeal and coherence.

The analyzed outcomes of the user study are shown in Fig. 7. The study participants ranked our approach as the closest to the texture exemplar ($\mu \pm \sigma = 64.3 \pm 26.9\%$ of the participants), followed by the approach of Gutierrez et al. ($\mu \pm \sigma = 17.6 \pm 16.5\%$) and Höllein et al. ($\mu \pm \sigma = 17.6 \pm 29.5\%$), and last by the approach of Mordvintsev et al. ($\mu \pm \sigma = 0.5 \pm 1.3\%$). The results are statistically significant ($F = 8.44196$; $p = .001028$), as shown with an ANOVA test followed by pairwise *t*-tests. The approach of Mordvintsev et al. was judged as the least similar to the texture exemplar ($\mu \pm \sigma = 57.6 \pm 33.3\%$ of the participants), followed by the approach of Höllein et al. ($\mu \pm \sigma = 35.7 \pm 35.1\%$), and Gutierrez et al. ($\mu \pm \sigma = 4.8 \pm 6.6\%$), and last by ours ($\mu \pm \sigma = 1.9 \pm 2.6\%$). The results are statistically significant ($F = 6.2159$; $p = .004376$), as shown with an ANOVA test followed by pairwise *t*-tests. In terms of visual appeal and coherence, the overall preferred approach is ours as indicated also visually in the plots of Fig. 7. The distribution of the ratings of our approach differs statistically significantly from the other three approaches in coherence ($F = 15.08026$; $p = .000037$) and visual appeal ($F = 13.77907$, $p = .000065$), as shown with ANOVA tests followed by pairwise *t*-tests. To sum up, according to our study participants, our approach outperforms qualitatively the other three methods in all investigated aspects.

**Speed and Memory Comparison.** We measured the training times and peak VRAM usage during the training phase of all approaches. Recall that our approach only considers the texels that are being used by a given UV unwrapping, hence for our approach we give the measurements for each mesh together with the percentage of used texels. The measurements in Table 1 indicate that our training times are comparable to the other approaches, sometimes even surpassing them depending on the fraction of used UV space. Even though we currently require more memory than Gutierrez et al. and Höllein et al., this can be alleviated with a better implementation, however, we would still need to store the neighborhoods for con-volution and pooling. On the positive side, though, even low-end GPUs have enough memory to be able to utilize our method. Both training times and memory scale with the used UV space, but there are small deviations from a linear scaling, which could be caused by different sizes of pooling groups. Finally, in Table 2 we show the precomputation times for the neighborhoods during convolution and pooling, which are an order of magnitude smaller than the training times.

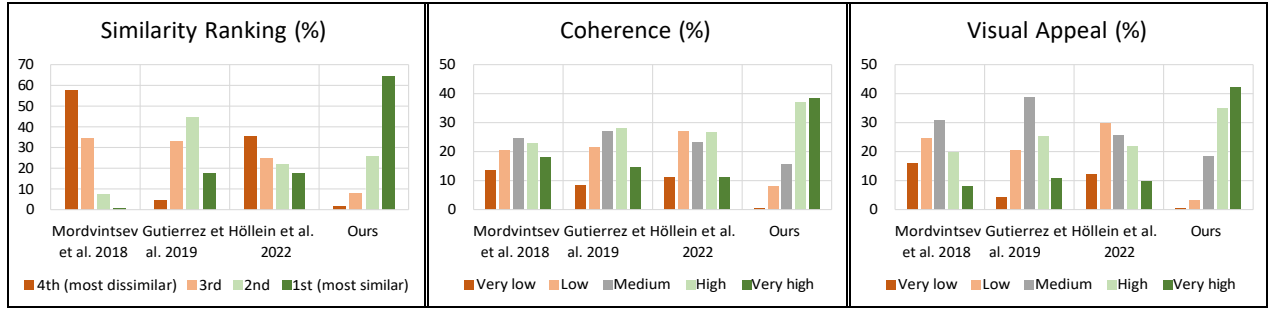

**Figure 7:** *Perceived similarity to the 2D exemplar, coherence, and visual appeal of our approach vs. Gutierrez et al. [GRGH19a], Mordvintsev et al. [MPSO18a], and Höllein et al. [HJN22a] in a user study with 30 participants.*

**Extension to Other Tasks.** Although we have heavily showcased our approach within the context of texture synthesis for single objects, there is a possibility of extension to a broader variety of tasks. This includes, for instance, style transfer and texture synthesis for whole scenes, but also segmentation or classification—similar to the previous work of Li et al. [LLZ*19]. We have not investigated the latter, but we showcase a few initial results of style transferring (Fig. 8) and stylization of whole scenes (Fig. 9). For the style transfer task, we have used an additional content texture that represents ambient occlusion and also an RGB texture [TL22]. For the former, we stylize with *The Great Wave* texture and the newspaper (used also by Mordvintsev et al. [MPSO18a]). We use the same approach as Gatys et al. [GEB15a], i.e., we match the Gram matrices of the generated texture with those of the style image and match the values of feature layers of both the generated texture and the original content texture. We use the output of the following layers to compute the Gram matrices: *block*1_*conv*1, *block*2_*conv*1, *block*3_*conv*1, *block*4_*conv*1, *block*5_*conv*1 which are weighted the same. We use the output of *block*4_*conv*2 to compute the content difference, which we multiply by 1000.

In these preliminary examples, we observe that our approach performs reasonably also for style transfer, despite not being explicitly designed for it. The style textures are applied in a manner that respects the geometry of the underlying mesh and the input texture. Observe the inner cavity of the bunny ear, and the creases under its neck or between its legs in Fig. 8 (a: top row); and compare them with the respective renderings provided in Fig. 10. The same effect can be noticed on the chest and the clothing of the Happy Buddha (Fig. 8, a: bottom row vs. Fig. 10). For the RGB texture, notice the contours around the eyes and nose of the bunny (Fig. 8 (b)). The same meshes and textures have been employed, with the sole addition of the content texture for the style transfer task.

For the whole scene texture synthesis task, we provide a comparison of our results with the approach of Höllein et al. in Fig. 9. The mesh was reconstructed from real-world data, therefore, it contains several holes and is noisy. This poses a challenge for our approach, as it favors surfaces that do not have a boundary. Furthermore, our method creates a rapidly changing tangent field in the noisy areas, which causes a loss of local context. Hence, the synthesis process is not able to create texture patches resembling the style exemplar. Oppositely, the approach of Höllein et al. does not have this problem. Yet, it struggles in a few areas close to the windows, which are at the bottom left part of the scene in Fig. 9. Those parts have not been sufficiently captured by the camera and, after rendering, the thin geometry is surrounded by a black background.

**Table 1:** *Training times (t) and peak VRAM usage for the 3D texture generation. For the render-based approaches of Mordvintsev et al. and Höllein et al., we used the bunny from the Stanford 3D scanning repository for the training. For the volume-based approach of Gutierrez et al., the measurements below refer to training with the entire volume. For our approach, we show results for different meshes. All measurements were done for a 1024 × 1024 marble texture (unless specified otherwise), and the outcomes are depicted in Figure 6. Note that the used UV space is not relevant for the measurements and is, thus, not indicated in the table.*

|  | t (m:s) | VRAM (MB) | Used UV |
|---|---|---|---|
| Gutierrez et al. | 59:43 | 564 | – |
| Mordvintsev et al. | 43:32 | 6122 | – |
| Höllein et al. | 40:48 | 711 | – |
| Ours – Bunny | 47:17 | 2422 | 63.33% |
| Ours – Bunny (512×512) | 12:39 | 737 | 64.53% |
| Ours – Bunny (256×256) | 3:51 | 291 | 66.84% |
| Ours – Armadillo | 43:53 | 2314 | 64.21% |
| Ours – Buddha | 49:46 | 2309 | 62.49% |
| Ours – Dragon | 38:49 | 2397 | 68.27% |
| Ours – Mother | 49:58 | 1933 | 52.61% |
| Ours – Snail | 36:07 | 1848 | 48.23% |
| Ours – Teapot | 53:24 | 2687 | 69.14% |

**Table 2:** *Precomputation times (t) of our approach using various meshes. A 1024 × 1024 texture (unless specified otherwise) is used for the 6 texel layers needed for VGG-19 with 5 pooling layers.*

|  | t (m:s) | Used UV | Triangles |
|---|---|---|---|
| Bunny | 1:44 | 63.33% | 5002 |
| Bunny (512×512) | 0:42 | 64.53% | 5002 |
| Bunny (256×256) | 0:21 | 66.84% | 5002 |
| Armadillo | 2:30 | 64.21% | 212574 |
| Buddha | 3:23 | 62.49% | 108770 |
| Dragon | 2:45 | 68.27% | 217853 |
| Mother | 1:33 | 52.61% | 56512 |
| Snail | 2:13 | 48.23% | 574 |
| Teapot | 3:32 | 69.14% | 6320 |

## 5. Limitations

Being inspired by the approach of Gatys et al., unavoidably our approach faces similar limitations. As discussed also in Sec. 4, we have high computational costs, especially in the optimization step. Furthermore, although our approach has reasonable control over the geometry and topology of the underlying mesh and produces visually appealing results as demonstrated in Sec. 4.1, we do not always have fine-grained control over the specific features or elements we want to transfer or retain in the synthesized texture. High-frequency changes in the tangent field may cause certain artifacts, e.g., colored spots that do not match the style texture. This is, for instance, visible in the examples with the plants texture (Fig. 4).

Our examples showcased that we are overall effective when applying a diverse set of textures or patterns—not only artistic styles but also natural textures and complex geometric patterns (Fig. 10). Yet, with a more robust architecture, we might be able to preserve better the texture structures or patterns. As expected, the algorithm's performance and the quality of the synthesized texture can be sensitive to several hyperparameters, requiring manual tuning and experimentation to achieve satisfactory results (Fig. 4 and 5). Moreover, uneven sampling and pooling may introduce information loss due to bias towards dominant features or regions in the data, as well as distortions or misalignments in the spatial relationships between features, impacting subsequent tasks that rely on accurate spatial information.

Lastly, our convolution definition in Eq. 2 assumes that the displacements along the geodesic paths are localized to a small neighborhood around the points. However, for network architectures with a large number of layers and, more importantly, several pooling operations, the covered patch might span large portions of the manifold breaking the assumption of locality.

## 6. Conclusion and Future Work

We have presented an example-based approach for texture synthesis for textured mesh objects. Our method uses a modification of the well-tested approach for style transfer of Gatys et al. [GEB15a], where the underlying data representation—instead of being a flat 2D plane—is the curved 3D surface of a given mesh. In this way, our approach takes into consideration the topology and geometry of the mesh in a manner superior to the previously proposed approaches. We showed that our approach works well for a variety of meshes with different styles. Our approach minimizes artifacts of existing learning-based methods, by being seamless and taking into account the local topology. Our method also minimizes feature bleeding across the Euclidean space.

In our future work, we will investigate 2D convolutional networks resistant to domain change from flat images to curved surfaces. It would also be interesting to modify the underlying mesh geometry to capture both visual style and 3D shape, similar to Hertz et al. [HHGCO20]. Another future direction could investigate other architectures, beyond VGG and with other loss functions. Additionally, our work could provide further insights into repurposing image-trained neural networks for general tasks with different local structures. Finally, in a future evaluation, it would be interesting to extend our approach to other tasks, e.g., segmentation or classification, and to compare our method to more recent NeRF-based stylization approaches [CYL*22, HHY*22, ZKB*22, LZC*23] and text-based stylization approaches [MBOL*21, KXBP22, RMA*23, MZS*23].

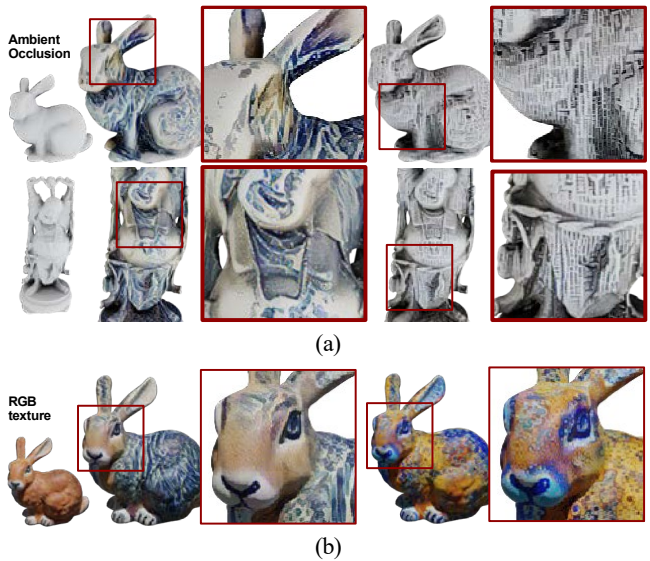

**Figure 8:** *(a) Two examples of style transfer with our approach with a texture that represents ambient occlusion. Top row: The input is the* Stanford bunny. *It is stylized with* The Great Wave off Kanagawa *(left) and a newspaper texture (right). Bottom row: The input is the* Happy Buddha, *stylized with the same two textures as the previous case. (b) Style transfer with two different styles (left:* The Great Wave off Kanagawa*, right:* Mandelbrot*) on the* Stanford bunny *with an RGB content texture [TL22].*

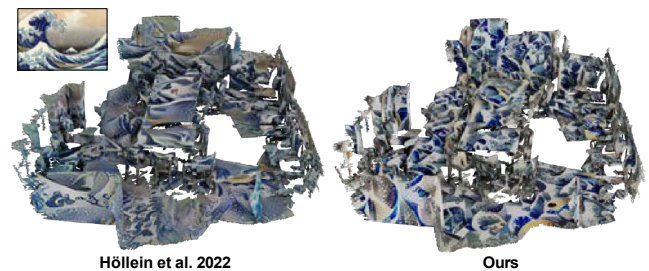

**Figure 9:** *Stylization for scene 0291_00 from the ScanNet dataset [DCS*17] achieved with the approach of Höllein et al. [HJN22a] and ours.*

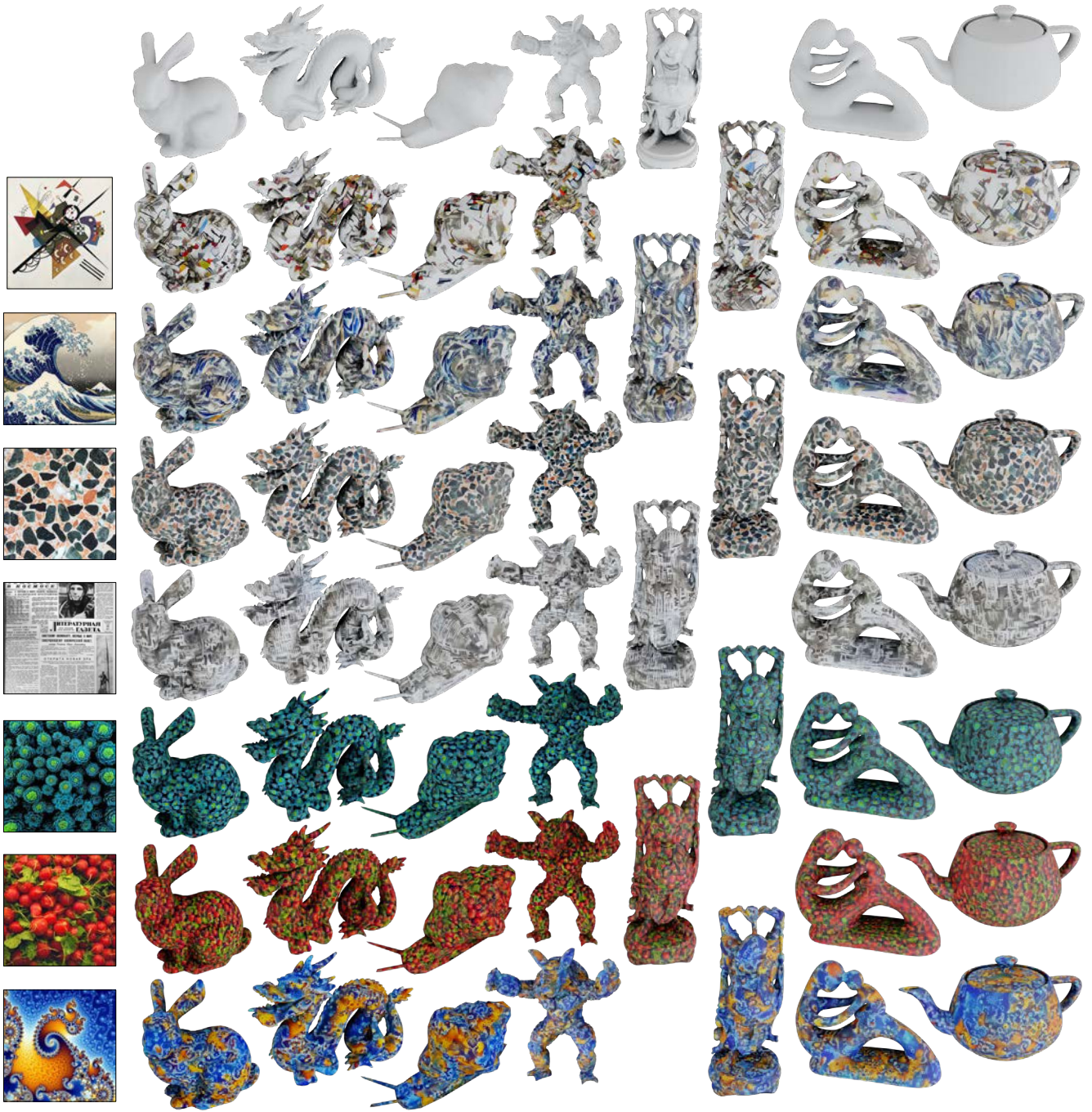

**Figure 10:** *Our method applied on seven meshes and seven textures with a diverse set of stimuli. The meshes include the bunny, the dragon, the armadillo, and the Happy Buddha from the Stanford 3D scanning repository, a snail mesh created in Blender, the Mother and Child by Brian Weston (CC BY-SA), and the Utah teapot. The textures include (from top to bottom) two artistic styles—namely, Kandinsky's on White II and The Great Wave off Kanagawa, an isotropic marble texture similar to those used by Gutierrez et al. [GRGH19a], the newspaper texture used by Mordvintsev et al. [MPSO18a], a high-frequency abstract texture containing succulent plants obtained from GitHub, the radishes texture also used by Gatys et al. [GEB15b] and previously by Portilla and Simoncelli [PS00], and the cropped Mandelbrot texture created by Wolfgang Beyer (CC BY-SA).*